\documentclass[a4paper,fleqn]{cas_sc}

\usepackage[numbers]{natbib}
\usepackage{xcolor}
\newcommand{\red}[1]{\textcolor{red}{#1}}

\def\tsc#1{\csdef{#1}{\textsc{\lowercase{#1}}\xspace}}
\tsc{WGM}
\tsc{QE}
\tsc{EP}
\tsc{PMS}
\tsc{BEC}
\tsc{DE}

\begin{document}
\let\WriteBookmarks\relax
\def\floatpagepagefraction{1}
\def\textpagefraction{.001}
\shorttitle{Evolutionary Routing in Complex Networks}
\shortauthors{F. Dilisante et~al.}

\title [mode = title]{When one protocol fits none: Self-organized network routing through evolutionary game dynamics}

\author[1,2]{Francesca Dilisante}[orcid=0009-0000-9419-4587]
\credit{Data curation, Methodology, Software, Writing - Original draft preparation}

\author[1,2]{Pablo Gallarta-Sáenz}[orcid=0009-0001-1681-9569]
\credit{Data curation, Methodology, Software, Writing - Original draft preparation}

\author[3]{Luciano Stucchi}[orcid=0000-0002-2823-0778]
\credit{Methodology, Writing - Original draft preparation}

\author[4,5]{Sandro Meloni}[orcid=0000-0001-6202-3302]
\credit{Conceptualization of this study, Methodology}

\author[1,2]{Jesús Gómez-Gardeñes}[orcid=0000-0001-5204-1937]
\cortext[cor1]{Corresponding author}
\credit{Conceptualization of this study, Methodology, Software, Writing - Original draft preparation}

\affiliation[1]{organization={Department of Condensed Matter Physics, University of Zaragoza},
                city={Zaragoza},
                postcode={50006},
                country={Spain}}
\affiliation[2]{organization={GOTHAM lab, Institute for Biocomputation and Physics of Complex Systems, University of Zaragoza},
                city={Zaragoza},
                postcode={50018},
                country={Spain}}
\affiliation[3]{organization={Universidad del Pacífico},
                city={Lima},
                postcode={15072},  
                country={Perú}}
\affiliation[4]{organization={Institute for Cross-Disciplinary Physics and Complex Systems (IFISC), CSIC-UIB},
                city={Palma de Mallorca},
                postcode={07122},
                country={Spain}}

\affiliation[5]{organization={Centro Studi e Ricerche ''Enrico Fermi'' (CREF), CNR},
                city={Rome},
                postcode={00184},
                country={Italy}}

\begin{abstract}
Packet routing on scale-free networks faces a fundamental trade-off: shortest-path routing is efficient at low demand but funnels traffic through hubs and jams early, whereas congestion-aware routing postpones jamming at the price of a sharper collapse. Since neither paradigm dominates across the full range of traffic load, here we ask whether the appropriate balance can emerge endogenously rather than being imposed by design. To answer this, we recast adaptive packet routing on networks as an evolutionary game letting a heterogeneous population of strategies compete for prevalence under selection pressure generated by their own performance. We study this competition under two formalisms (strategy anchored to the packet or to the generating node), global and local update rules, and two payoff metrics. Across every implementation the evolutionary dynamics yield the same outcome: the jamming transition is delayed relative to shortest-path routing while the violent collapse of fixed congestion-aware routing is avoided. This improvement emerges spontaneously, without centralized coordination or global information. Crucially, under local update rules, the node-level volatility of strategy choices peaks sharply at the transition, furnishing a purely local early-warning signal of imminent jamming that requires no global monitoring.
\end{abstract}


\maketitle

\section{Introduction}
\label{sec:intro}

Many of the infrastructures that sustain modern societies, from the Internet and transportation grids to the metabolic and neural circuits of living organisms, can be cast as networks~\cite{albert2002statistical,newman2003structure,boccaletti2006complex} through which discrete units of information, matter, or energy are routed between their corresponding sources and targets~\cite{masuda2017random}. A defining feature of these networked systems is their heterogeneous, often scale-free~\cite{BA1999Science, BA2000Nature}, organization in which a few highly connected hubs coexist with a large majority of poorly connected nodes. This heterogeneity makes transport remarkably efficient at low demand, but it also concentrates traffic on the hubs, so that the same backbone that accelerates delivery becomes the bottleneck that precipitates congestion as load grows.
\smallskip

The interplay between topology, demand, and routing rule is expressed most sharply in the jamming transition exhibited by packet-routing models on complex networks \cite{Sawatari1998PRE, arenas2001communication, guimera2002optimal, Echenique2004PRE,zhao2005onset, Echenique2005EPL,danila2006optimal,toroczkai2004jamming}. As the rate at which packets are injected increases, the network passes from a free-flow phase, in which packets are delivered as fast as they are created, to a congested phase, in which undelivered packets accumulate without bound. The critical rate separating the two is a primary measure of performance: the larger the better. Crucially, the value of the critical load and the very nature of the transition depend on how packets are forwarded. Shortest-path (SP) routing minimizes travel distance but funnels traffic through the hubs, jamming the network at low values of flow through a smooth, second-order-like, transition. However, congestion-aware (CA) routing, which steers packets away from loaded nodes, postpones the onset of jamming to a higher load, but pays for it with a sharper, more abrupt collapse once capacity is exceeded \cite{Echenique2004PRE, Echenique2005EPL}. These two paradigms thus trace out qualitatively different congestion diagrams and, decisively for what follows, neither dominates the other across the whole range of demand: the strategy that is optimal at low load is not the one that is optimal near saturation.
\smallskip

The absence of a unique best protocol has motivated adaptive routing schemes that seek to combine the virtues of both extremes. Strategies based on local {\em empathy}, in which each node accounts for the congestion it imposes on its neighbors, minimize global congestion using only locally available information \cite{Meloni2010PRE}, while adaptive and reciprocal interaction rules keep networks congestion-free by reshaping the flow or the topology in response to traffic \cite{Gavalda2012PRE}. These works show that hybrid approaches provide the adequate flexibility between SP-like and CA-like behavior, yet they prescribe in advance the mechanism by which that flexibility is exercised. A complementary question, which we tackle here, is whether the appropriate balance can instead emerge endogenously, with the routing logic of the network selected by its own performance rather than imposed by design.
\smallskip

Posing the problem in these terms naturally casts it as an evolutionary game. When no single strategy is optimal across the full range of conditions, and the benefit of a given behavior depends on how many agents already adopt it, the best course of action cannot be fixed in advance: it can only be defined self-consistently, through the competition among the strategies themselves. Evolutionary game theory has proven especially powerful for precisely this class of problems~\cite{Wang2016PhysRep}, in which individual incentives and collective outcomes are misaligned and no behavior is universally optimal. A paradigmatic example is the vaccination dilemma~\cite{BauchEarn2004PNAS,BauchGalvaniEarn2003PNAS} on networks, where agents choose whether or not to vaccinate and the more advantageous choice depends on the prevailing infectivity and on the decisions of others \cite{Cardillo2013PRE}. Subsequent work has shown that the interplay between the individual cost and the collective benefit of protection produces nontrivial uptake patterns \cite{Steinegger2018PRE}, that risk-driven prophylaxis can sustain oscillations in disease prevalence \cite{Steinegger2020PRR}, and that heterogeneous perceptions of risk reshape the emergence of protective behavior \cite{Khanjanianpak2022PTRSA}. In all of these settings the macroscopic state of the system arises self-consistently from decentralized, self-interested choices rather than from any globally optimal prescription.
\smallskip

Building on this perspective, we recast adaptive packet routing on scale-free networks as an evolutionary game. Rather than fixing a single forwarding rule, we let a heterogeneous population of routing strategies compete for prevalence under selection pressure generated by their own differential performance. We analyze this competition under two formalisms that differ in where strategy is anchored (the individual packet versus the generating node), under both global (mean-field) and strictly local update rules, and under two payoff metrics that emphasize complementary facets of routing efficiency. Across all of these implementations the evolutionary dynamics yield the same qualitative outcome: the jamming transition is delayed relative to fixed SP routing, while the violent collapse characteristic of fixed CA routing is avoided, and this improvement emerges spontaneously, without centralized coordination or global information. We further show that, under local update rules, the node-level volatility of strategy choices furnishes a purely local early-warning signal of the impending jamming transition. 
\smallskip

\section{Model and theoretical framework}
\label{sec:2}

The framework is organized in three layers. The first defines the network substrate and microscopic traffic dynamics: a scale-free topology on which packets are injected stochastically, queued under a First-In-First-Out (FIFO) discipline, and forwarded by a congestion-aware rule that couples topological and traffic information. The second superimposes evolutionary dynamics, treating the congestion-awareness parameter $h$ as a heritable strategy under selection. The third specifies the fitness landscape through two payoff metrics that quantify routing performance.
\smallskip

\subsection{Network model and traffic dynamics}
\label{sec:model_network}

The network substrate consists of $N$ nodes with a scale-free architecture \cite{BA1999Science, BA2000Nature}, with degree distribution $P(k)\sim k^{-\gamma}$ and $\gamma \approx 3$, consistent with real communication infrastructures~\cite{BA1999Science}. The routing dynamics are built on the congestion-aware model of Echenique et al.~\cite{Echenique2004PRE, Echenique2005EPL}. At each discrete time step, every node generates a packet with probability $p$ ---the primary control parameter--- and assigns it a destination drawn uniformly among the remaining $N-1$ nodes. Packets queue in each node's local buffer under a FIFO discipline: only the head-of-line packet is eligible for forwarding per time step, a serialization that proxies the finite bandwidth of physical routers.
\smallskip

The microscopic state at time $t$ is given by the queue lengths $\{ q_i(t)\}_{i=1,\ldots,N}$, whose sum defines the global load $Q(t)$,
\begin{equation}
    Q(t) = \sum_{i=1}^N q_i(t).
\end{equation}
The long-run behavior of $Q(t)$ diagnoses the dynamical state: a free-flow phase where $Q(t)$ fluctuates around a finite value, and a congested phase where it grows without bound \cite{Echenique2005EPL}. We quantify this through an order parameter $\rho$, the normalized stationary rate of change of the total load over a window $\Delta t$,
\begin{equation}
    \rho(t)=\lim_{t\rightarrow \infty} \dfrac{Q(t+\Delta t)-Q(t)}{pN\Delta t},
\end{equation}
\noindent where the normalization $pN\Delta t$ is the expected number of packets injected during the interval, rendering $\rho$ dimensionless and bounded. In free flow, packets are cleared at their creation rate and $\rho\to 0$; in congestion, they accumulate and $\rho$ takes a non-zero value. The system thus undergoes a transition at a critical probability $p_c$ that, depending on topology and routing strategy \cite{Echenique2004PRE, Sawatari1998PRE}, serves as a primary performance descriptor throughout this work.
\smallskip

The transition is sensitive to the forwarding rule, and it is here that adaptive congestion-aware models depart from classical ones \cite{Echenique2004PRE, Echenique2005EPL, Feng2010CPB, Zhang2007PLA}. Rather than a static shortest-path (SP) protocol ---which concentrates load on hubs and jams at low $p$ \cite{Echenique2004PRE, Sawatari1998PRE}--- the congestion-aware (CA) rule uses a local effective distance that couples topological information to the instantaneous traffic state, letting each node balance path efficiency and load distribution in real time (see Fig.~\ref{fig:0}(a)).
\smallskip

Formally, when node $k$ prepares to send its head-of-line packet toward destination $j$, it scores each of its neighbors $i \in \mathcal{N}(k)$ according to:
\begin{equation}
    d_{ij}^{\mathrm{eff}} = h\cdot d_{ij} + (1 - h)\cdot q_i(t),
    \label{eff_dist}
\end{equation}
\noindent where $d_{ij}$ is the topological shortest-path distance from neighbor $i$ to destination $j$, and $q_i(t)$ its queue length. The packet is forwarded to the neighbor $i^*$ minimizing $d_{ij}^{\mathrm{eff}}(t)$. The parameter $h \in [0,1]$ acts as a tunable control parameter that governs the routing strategy: in the limit $h\to 1$, the effective distance reduces to $d_{ij}$, recovering pure shortest-path routing; in the limit $h \to 0$, the score function is governed entirely by $q_i(t)$, encoding a purely congestion-avoidance strategy that directs traffic away from loaded nodes regardless of path length. Intermediate values of $h$ define a continuously adjustable family of strategies that trade off path efficiency against avoidance of long buffer wait times. It is precisely within this family that the evolutionary dynamics introduced in the following section will operate.
\smallskip

\subsection{Evolutionary game theory (EGT) framework}
\smallskip

The integration of evolutionary game theory is motivated by the tension between the two classical paradigms. On one hand, SP routing is topologically optimal at low traffic levels but undergoes a second-order transition at a low critical rate $p_c^{\mathrm{SP}}$, as load concentrates irreversibly on high betweenness nodes such as hubs \cite{Echenique2004PRE, Echenique2005EPL}. On the other hand, CA routing bypass this problem by redistributing load away from hubs, delaying jamming to $p_c^{\mathrm{CA}} > p_c^{\mathrm{SP}}$, but at the cost of a sharper, first-order-like collapse once capacity is saturated \cite{Echenique2004PRE, Echenique2005EPL}. Research on this phenomena pinpoint that this explosive behavior is due to the myopic (short range) nature of the Eq.~\ref{eff_dist}, since globally informed CA variants recover second-order behavior \cite{Feng2010CPB, Zhang2007PLA}.
\smallskip

This complementarity, with no single static protocol capturing both advantages, motivates an evolutionary competition among a heterogeneous population of $m$ discrete strategies, with $h$ uniformly sampled over $[0,1]$, thus constituting a multi-strategic environment. Rather than fixing a protocol, the partition of routing strategies adapts in time through a sequence of discrete updates, the \textit{generations}, each comparing the strategies' performance and reallocating their prevalence accordingly. A we explain below, this adaptation is intrinsically endogenous, as it operates under selection pressure from the differential performance of competing strategies, driving the population toward evolutionarily stable distributions that exploit the complementarity between path efficiency and congestion avoidance.
\smallskip

To tackle the evolutionary dynamics of routing strategies we analyze two paradigms that differ in where strategy is anchored: (i) the \textit{packet-centric} formalism, where strategy is an intrinsic attribute assigned to each packet at its creation (so selection acts on the packet flow), and (ii) the \textit{node-parent} formalism, where strategy is anchored to the generating node (so that it applies its $h$ to all packets it produces). Together, these two paradigms span a natural spectrum of assumptions about the nature of strategy allocation mechanisms in distributed routing systems.
\smallskip

\subsubsection{Packet-centric formalism}
\smallskip

\begin{figure}[]
    \centering
    \includegraphics[width=\linewidth]{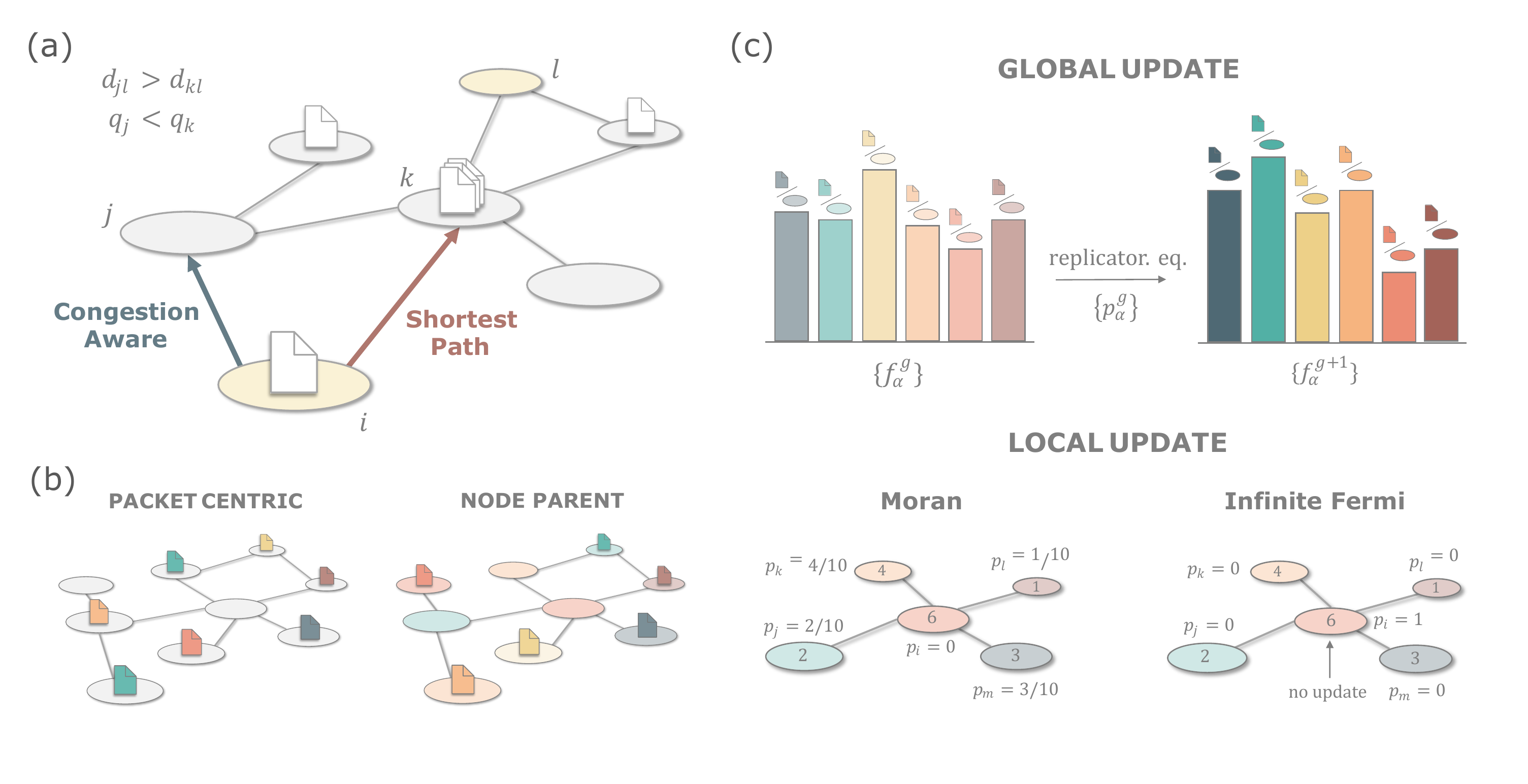}
    \caption{\textbf{Schematic representation of the main elements of the framework.} Panel (a) illustrates the SP (red) versus CA (blue) routing paradigms. Under SP routing, the packet at the highlighted node, $i$, is forwarded to the neighbor that minimizes topological distance to the destination, $l$, regardless of queue lengths. Under CA routing, heavily loaded neighbors are avoided in favor of less congested ones, even at the cost of a longer topological path. Panel (b) contrasts the packet-centric and node-parent formalisms. In the node-parent formalism, each node is assigned a strategy that it applies to every packet it generates. In the packet-centric formalism, each newly created packet is assigned a strategy independently, drawn from the current strategy distribution, irrespective of its origin node. In both cases, the assigned strategy is fixed for the packet's lifetime. Panel (c) illustrates the update rules considered in this work. The global rule (replicator equation), used in both the packet-centric and node-parent formalisms, updates the strategy distribution at each evolutionary time step in proportion to each strategy's payoff relative to the population average. The local rules (Moran and Infinite Fermi), considered only within the node-parent formalism, update strategies through pairwise neighborhood interactions; the panel illustrates how the same local configuration can favor different strategies depending on which rule is applied. Numbers on the nodes refer to node payoff while $p_j$ refers to the probability that node $i$ chooses to copy node $j$'s strategy.}
    \label{fig:0}
\end{figure}

In the packet-centric formalism, strategy is an immutable attribute assigned to each packet at creation in proportion to the current strategy distribution (see Fig.~\ref{fig:0}(b)). Thus, by starting from a uniform distribution $f_\alpha(n=1) = 1/m$, each generation unfolds in two stages: first, traffic is simulated under $\{f_\alpha(n)\}$ until $Q(t)$ and $\rho$ reach a steady state and, second, the performance of each strategy is evaluated through the payoff functions that will be defined in Section \ref{section:payoffs}. 
\smallskip

The distribution evolves according to the discrete-time replicator equation (see Fig.~\ref{fig:0}(c)), whose simplest version~\cite{Chakraborty2018Chaos} reads:
\begin{equation}\label{eqn:replicator}
f_\alpha(n+1) = f_\alpha(n)\cdot\frac{p_\alpha(n)}{\langle p_\alpha(n) \rangle},
\end{equation}
where $p_\alpha(n)$ is the payoff of strategy $\alpha$ and $\langle p_\alpha(n) \rangle = \sum_\beta f_\beta(n)\, p_\beta(n)$ the mean payoff. Thus, the former expression embodies the core logic of natural selection: strategies whose payoff exceeds the population average increase their fractional representation in the subsequent generation, while below-average strategies are progressively displaced. Besides, the replicator equation preserves the total mass, $\sum_\alpha f_\alpha(n) = 1$. The iteration across successive generations drives the distribution to an evolutionarily stable fixed point.
\smallskip

\subsubsection{Node-parent formalism}\label{sec:NP_description}
\smallskip

Packet-centric formalism allows routing to vary on a per-packet basis within a node, which is convenient from an analytical standpoint but comes into conflict with real routers, which apply a single, permanent policy. To solve this problem, the node-parent formalism instead anchors strategy to the node, {\em i.e.}, each node $i$ has a fixed $h_i$ applied to every packet it creates (see Fig.~\ref{fig:0}(b)). Here adaptation can proceed through two update mechanisms with different informational demands: global or local. 
\smallskip

Under a \textit{global update} rule, each strategy's payoff aggregates performance across all nodes assigned to it, and frequencies update via the same replicator equation~(\ref{eqn:replicator}) (see Fig.~\ref{fig:0}(c)). Thus, the selection remains mean-field: all nodes update simultaneously from system-wide statistics, isolating the effect of the strategy-assignment locus (packet versus node) while holding the update rule fixed.
\smallskip

The global rule assumes every agent accesses network-wide performance, which is unrealistic since real routing agents observe only their own traffic and, at best, coarse signals from adjacent neighbors. This tension motivates \textit{local update} rules, under which adaptation proceeds solely from information within each node's immediate neighborhood, without any node tracking the global strategy distribution or payoff landscape.
\smallskip

Under a local update rule, adoption is governed by pairwise neighborhood interactions (see Fig.~\ref{fig:0}(c)). We consider two widely used rules \cite{Cuesta2009PRE, Szabo2007PR}: Moran \cite{Cuesta2009PRE, Moran1962} and Fermi \cite{SzaboPRE1998, SzaboAJP2005}. Under the Moran rule, node $i$ adopts the strategy of a neighbor $j \in \mathcal{N}(i)$ with probability proportional to $j$'s payoff relative to the neighborhood total, {\em i.e.}:
\begin{equation}
    \Pi_{i \rightarrow j} = \frac{P_j}{\sum_{k \in \mathcal{N}(i)} P_k}.
\end{equation}
Under \textit{Fermi} update rule, each node $i$ selects a neighbor $j \in \mathcal{N}(i)$ at random and adopts its routing strategy with a probability given by the Fermi distribution,
\begin{equation}
    \Pi_{i \rightarrow j} = \frac{1}{1+\exp{[-\beta\Delta_{ji}]}}.
\end{equation}
where $\Delta_{ij} = P_j - P_i$ is the signed payoff difference and $\beta \geq 0$ tunes the sensitivity to fitness differentials. At finite $\beta$ adoption is always positive, even when $P_j < P_i$, whereas as $\beta \to \infty$ the Fermi update rule becomes a step function, so $i$ copies $j$ if and only if $P_j > P_i$. This deterministic limit defines the \textit{Infinite Fermi} rule adopted here. This way, by removing stochastic exploration, we isolate the pure selection dynamics of local fitness comparisons.
\smallskip

The Infinite Fermi rule differs from Moran in two respects. First, the role model is chosen uniformly at random rather than by payoff-weighted sampling and, second, adoption is deterministic since imitation occurs only if the neighbor strictly outperforms the focal node. Moran, by contrast, can copy any neighbor with positive probability, allowing escape from local fitness minima. Infinite Fermi thus converges faster to locally dominant strategies but explores less strategic configurations.
\smallskip

In a nutshell, local rules mirror the absence of global information in distributed systems since node\textcolor{red}{s} do not need the system-wide distribution or mean payoff to make their update. In addition, as a byproduct, the resulting evolutionary dynamics are spatially structured, with successful strategies propagating along network topology.
\smallskip

\subsection{Payoff functions}\label{section:payoffs}
\smallskip

The fitness landscape governing the evolutionary adaptation of routing strategies is defined through two payoff metrics capturing complementary dimensions of routing performance. Using multiple metrics is deliberate: congestion avoidance, transit speed, and robustness to delivery failure are not generally optimized by the same strategy, so comparing evolutionary outcomes across metrics is itself informative.

\subsubsection*{Delivery-efficiency payoff}
\smallskip

The first metric quantifies a strategy's ability to keep the network in the free-flow regime. Formally, for the global update rule, it is defined as the complement of the strategy-resolved order parameter $\rho_\alpha$,
\begin{equation}
    P_\alpha^{\mathrm{DE}} = 1 - \rho_\alpha=\lim_{t\rightarrow\infty}\dfrac{Q_\alpha(t+\Delta t)-Q_\alpha(t)}{Np\Delta t f_\alpha},
\end{equation}
where $Q_\alpha(t)$ is the contribution to total network load from packets operating under strategy $\alpha$, and $f_\alpha$ is the current fraction of packets assigned to that strategy, so that $N p \Delta t \, f_\alpha$ represents the expected number of strategy $\alpha$ packets injected during the observation window. The quantity $\rho_\alpha \in [0, 1]$ measures the normalized net accumulation rate of undelivered packets under strategy $\alpha$: it vanishes when the strategy successfully clears packets at the rate they are injected, and approaches unity under complete congestion. Accordingly, the delivery-efficiency payoff $P_\alpha^{\mathrm{DE}}$ takes a maximum value of one in the free-flow phase and decreases monotonically as the strategy drives the system toward jamming, penalizing routing behaviors that contribute to systemic packet accumulation.
\smallskip

\subsubsection*{Global-efficiency payoff}
\smallskip

The second metric assesses transit efficiency with an explicit penalty for delivery failure. It captures instantaneous packet velocity so that a packet advancing every step contributes fully while one stuck in queues does it less. In addition we assign zero velocity to those packets undelivered at the end of the evaluation window. Under the global update rule, the metric reads:
\begin{equation}
    P_\alpha^{\mathrm{GE}} =\lim_{t\rightarrow\infty}\dfrac{\sum_{i\in\{\mathcal{D}_\alpha\}}\dfrac{d_i^r}{\tau_i}}{Np\Delta T f_\alpha}.
\end{equation}
where the sum runs over delivered packets only, and the denominator $Np\Delta T f_\alpha$ is the total number of packets generated under strategy $\alpha$ during the observation window, thus introducing the delivery penalty directly into the normalization.
\smallskip

The resolution at which payoff is computed depends on the update rule. Under the packet-centric formalism and the node-parent global rule (both governed by the replicator equation), $P_\alpha(n)$ is a strategy-level quantity, thus aggregating performance over all packets assigned to strategy $\alpha$ regardless of origin. Instead, the local rules (Moran and Infinite Fermi), operating through pairwise comparisons between neighboring nodes, require node-level resolution, {\em i.e.}, the payoff $P_i$ of node $i$ is computed only from the packets it has generated, pointing out the locality of information available to a real router.
\smallskip

With the three layers specified (network substrate, congestion-aware routing together with its evolutionary reinterpretation, and the payoff functions), the framework is ready for numerical exploration. The nonlinear coupling between queue dynamics, routing decisions, and the evolving strategy distribution yields a high-dimensional stochastic system with no closed-form solution, so its stationary and transient properties must be accessed computationally. This is intrinsic to the phenomena under study: the emergent routing efficiency is a collective behavior that resists reduction to any single component in isolation.
\smallskip

\section{Results}
\label{sec:3}
\smallskip

Once introduced the main ingredients of the evolutionary routing framework, in this section we examine the behavior of the system under progressively more realistic conditions: (i) the packet-centric formalism under both payoff metrics; (ii) the node-parent formalism, comparing both metrics and global (replicator) versus local (Moran and Infinite Fermi) update rules; and (iii) for the local rules, how node-level measurements can serve as early indicators of the jamming transition without global information. Throughout, we track the strategy distribution, identifying its fixed points and characterizing when coexistence/dominance emerges and how it shapes global congestion.
\smallskip

\subsection{Simulation setup}
\label{sec:setup}
\smallskip

To assess this evolutionary routing framework, simulations were conducted on a scale-free (SF) network generated via the Barabási-Albert model with $N = 1000$ nodes and mean degree $\langle k \rangle = 4$, consistent with Section~\ref{sec:model_network}. All observables and strategy distributions are averaged over 100 independent realizations.
\smallskip

Each evolutionary generation spans $T = 5 \times 10^4$ time steps, split into a thermalization period of $T_{\mathrm{eq}} = 10^4$ steps plus a measurement window of $T_{\mathrm{meas}} = 4 \times 10^4$ steps. Thermalization is essential to ensure that the payoff functions are evaluated under stationary traffic conditions, in which traffic relaxes toward a steady state and $\{q_i(t)\}$ and $Q(t)$ stabilize. After this period, payoffs and the order parameter $\rho$ are then averaged over the measurement window. We verified that $T_{\mathrm{eq}}$ suffices for $Q(t)$ to converge across the full range of $p$ and strategy configurations, and that $T_{\mathrm{meas}}$ is long enough to suppress fluctuations in the time-averaged payoffs.
\smallskip

\begin{figure}[]
\centering\includegraphics[width=\linewidth]{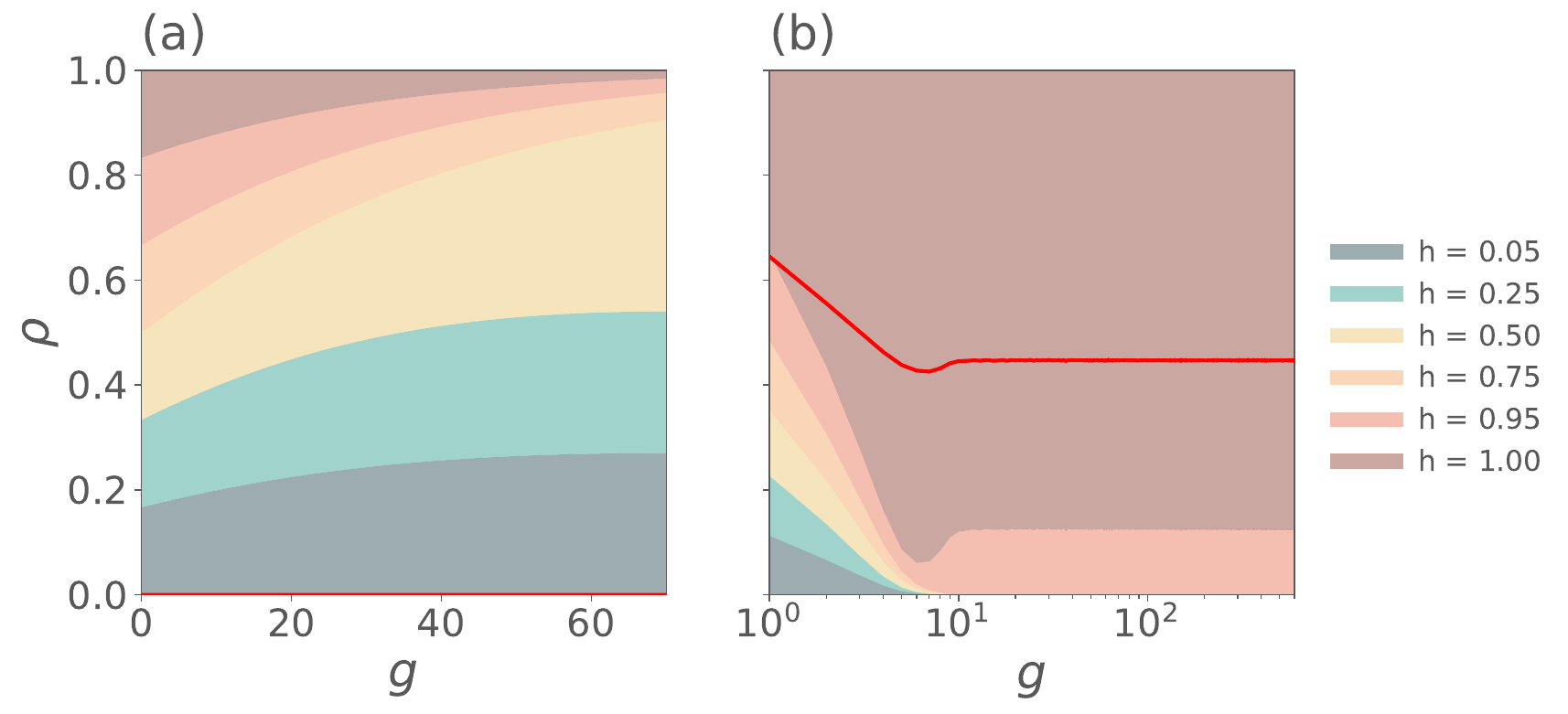}
\caption{\textbf{Strategy distribution evolutionary dynamics in the packet-centric formalism.} The panels illustrate the strategy distribution $f_\alpha$ (color map) as a function of the generation for two different values of packet injection probability $p$ using the global-efficiency payoff: (a) $p$=0.004, and (b) $p$=0.016. The red solid line indicates the evolutionarily determined order parameter $\rho$, representing the global network congestion. Results are averaged over 100 independent realizations on a scale-free network with $N=1000$ and $\langle k \rangle = 4$.}
\label{fig:1}
\end{figure}

Convergence to a stationary strategy distribution $\{f_\alpha^*\}$, whose composition depends on $p$, is assessed dynamically: stationarity is declared once $\{f_\alpha(g)\}$ becomes statistically indistinguishable from $\{f_\alpha(g-\Delta g)\}$ over a prescribed window $\Delta g$. A hard ceiling of $G = 600$ generations is imposed for tractability and justified \textit{a posteriori}, since for all parameter combinations $\{f_\alpha(g)\}$ reaches its stationary value well before this limit. Thus, this protocol ensures that all reported results correspond to evolutionarily stable configurations rather than to transient states that have not yet reached their asymptotic values. In Fig.~\ref{fig:1} we illustrate this for the packet-centric formalism under $P^{\mathrm{GE}}$ while equivalent convergence has been systematically verified for all payoff metrics, evolutionary paradigms, and parameter combinations presented in the main text, with the corresponding generational evolution plots included in the Supplementary Material.
\smallskip

\subsection{Packet-centric evolutionary dynamics}
\smallskip

The system's response to increasing traffic load $p$ under the packet-centric formalism is summarized in Fig.~\ref{fig:2}, which shows the stationary strategy distributions $\{f_\alpha\}$ and the corresponding order parameter $\rho$ for both payoff metrics. As a benchmark, we include the congestion profiles for two non-evolutionary scenarios: pure shortest-path routing ($h = 1$, dashed line) and a representative congestion-aware strategy ($h = 0.95$, dotted line) whose behavior under fixed conditions was discussed in Section~\ref{sec:model_network}.
\smallskip

\subsubsection{Delivery-efficiency payoff}
\smallskip

Under the delivery-efficiency payoff, which rewards minimizing packet accumulation, the evolutionary dynamics extend the free-flow regime, raising $p_c$ above the fixed SP baseline. For $p < p_c$ (Fig.~\ref{fig:2}(a)) all strategies coexist in a nearly uniform distribution (see Fig.~\ref{fig:SM_2} of Supplementary Material). In this low-traffic regime queues rarely form and, therefore, all strategies deliver successfully yielding a negligible selection pressure and leaving the strategies evolutionarily neutral.
\smallskip

This neutrality, nonetheless, conceals a structural benefit: even at low $p$, the presence of strategies with $h \neq 1$ redistributes traffic away from bottlenecks (high-degree hubs) thus reducing the load concentration that triggers jamming under pure SP. The population thus attains a higher effective $p_c$ than fixed SP without any coordination, purely through the strategic diversity sustained in the sub-critical regime.
\smallskip

Near $p_c$ the transition is qualitatively distinct: $\rho$ jumps in a first-order fashion but, importantly, only to values comparable to fixed SP at the same $p$, thus avoiding the sharper collapse of purely CA routing. As congestion sets in, strategies contributing disproportionately to the backlog are eliminated and the population converges toward near-SP routing, with only two survivors: a majority of pure SP packets ($h=1.0$) and a minority of moderately congestion-aware ones ($h=0.95$). This evolutionary hedging keeps the total load below a purely CA-driven system while retaining the delayed onset that diversity affords over fixed SP.
\smallskip

\subsubsection{Global-efficiency payoff}
\smallskip

The global-efficiency payoff jointly rewards transit speed and delivery reliability, assigning zero velocity to packets undelivered by the end of the window (Section~\ref{section:payoffs}), so it is sensitive both to how fast packets move and to whether they arrive.
\smallskip

\begin{figure}[]
\centering\includegraphics[width=\linewidth]{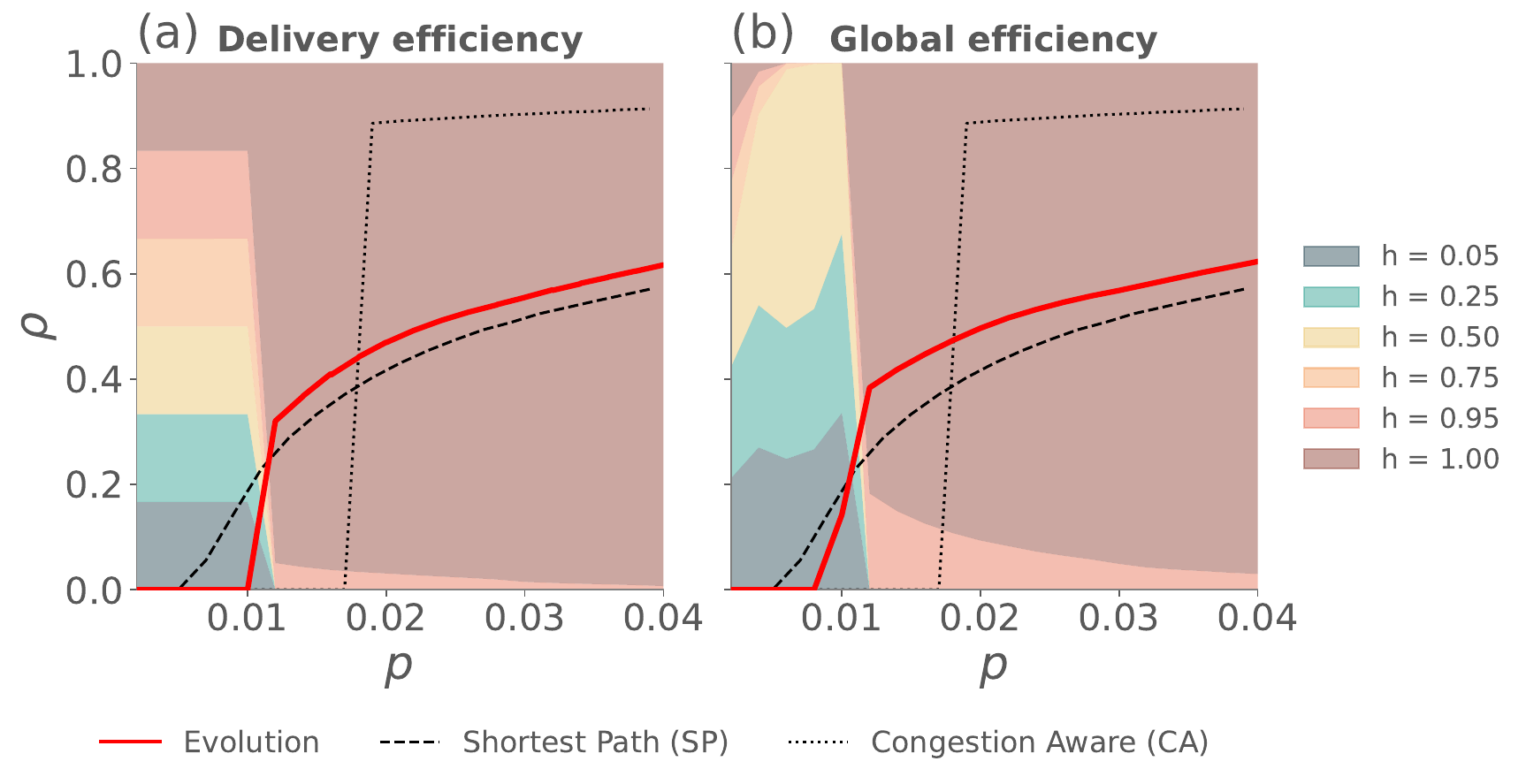}
\caption{\textbf{Stationary state of the evolutionary routing dynamics in the packet-centric formalism. }The panels illustrate the steady-state strategy distribution $f_\alpha$ (color map) as a function of the packet injection probability $p$ for two distinct fitness landscapes: delivery-efficiency payoff (left), and global-efficiency payoff (right). The red solid line indicates the evolutionarily determined order parameter $\rho$, representing the global network congestion. For comparison, the dashed and dotted lines provide the congestion baselines for static, non-evolutionary routing strategies: pure shortest-path ($h=1$) and congestion-aware ($h=0.95$), respectively. Results are averaged over 100 independent realizations on a scale-free network with $N=1000$ and $\langle k \rangle = 4$.}
\label{fig:2}
\end{figure}

The resulting dynamics, shown in Fig.~\ref{fig:2}(b), differ from the delivery-efficiency case mainly in the sub-critical regime. Since the metric rewards queue avoidance, congestion-aware strategies are favored even at very low $p$ and the distribution shifts toward lower $h$, yielding a slightly lower $p_c$.
\smallskip

Once the transition occurs, the global-efficiency metric qualitatively recovers the same robust post-critical behavior: the order parameter $\rho$ jumps to values comparable to those of fixed SP routing rather than collapsing to near-total congestion, and the system settles into a coexistence regime dominated by near-SP strategies. The global-efficiency payoff thus captures the transit speed advantages of congestion-aware routing in the free-flow regime while retaining the same protection against congestion collapse above $p_c$.
\smallskip

\subsection{Node-parent evolutionary dynamics}
\smallskip

The packet-centric formalism has shown that evolutionary dynamics can extend the free-flow regime and soften the jamming transition, but it does so under the assumption that strategy varies packet-by-packet within a node. We now turn to the node-parent formalism, where strategy is a persistent node property, and ask whether these outcomes survive this stricter constraint. Beyond the change in assignment, this formalism also opens the framework to local update rules (Moran and Infinite Fermi) alongside the global replicator, thus letting us probe how network structure and the locality of information shape the process.
\smallskip

\begin{figure}[]
\centering\includegraphics[width=\linewidth]{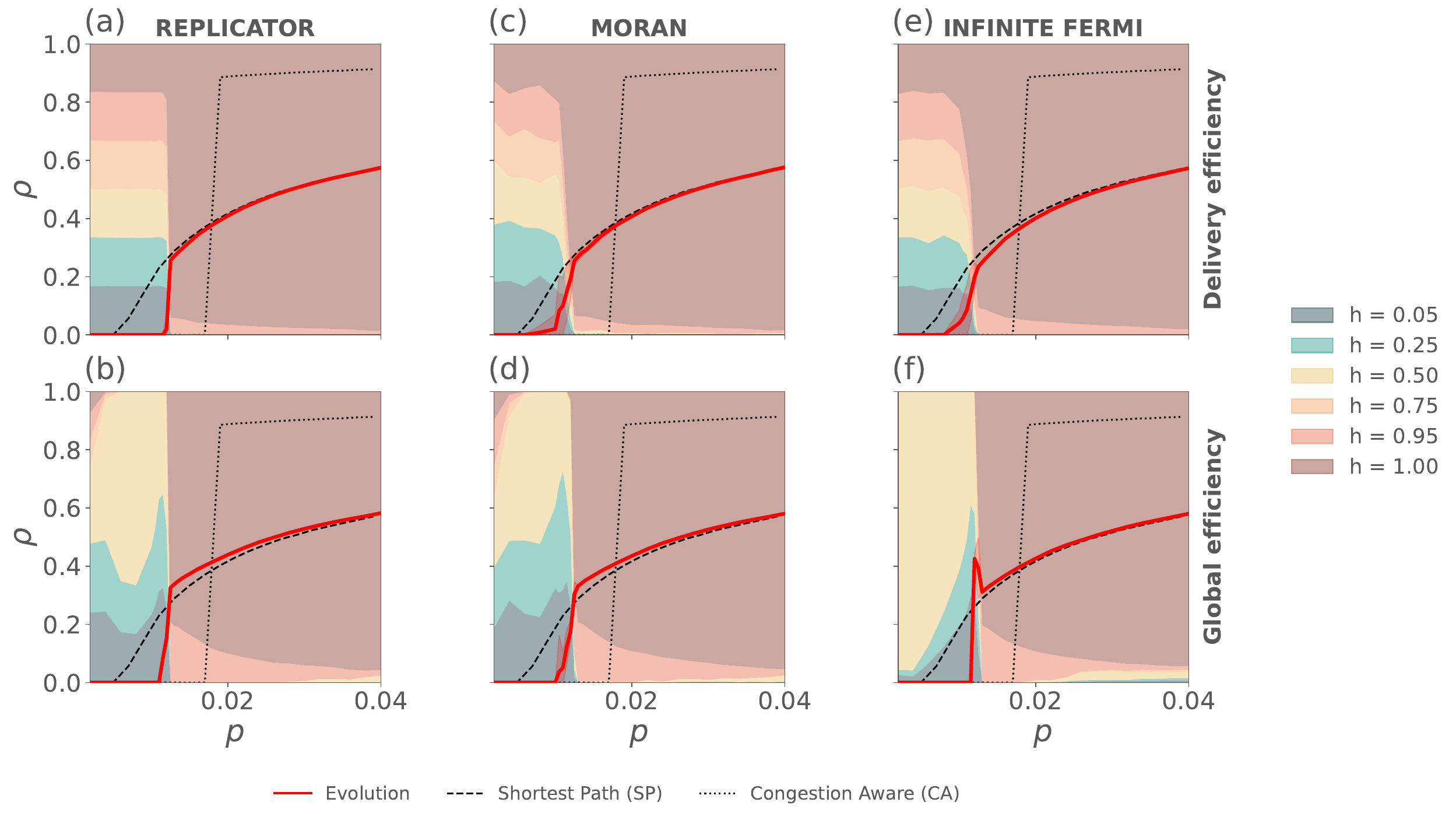}
\caption{\textbf{Stationary state of the evolutionary routing dynamics in the node-parent formalism.} The panels illustrate the steady-state strategy distribution $f_\alpha$ (color map) as a function of the packet injection probability $p$ for 6 distinct combinations of payoffs and update rules: (a) delivery-efficiency payoff under Replicator update rule, (b) global-efficiency payoff under Replicator update rule, (c) delivery-efficiency payoff under Moran update rule, (d) global-efficiency payoff under Moran update rule, (e) delivery-efficiency payoff under Infinite Fermi update rule and (f) global-efficiency payoff under Infinite Fermi update rule. The red solid line indicates the evolutionarily determined order parameter $\rho$, representing the global network congestion. For comparison, the dashed and dotted lines provide the congestion baselines for static, non-evolutionary routing strategies: pure shortest-path ($h=1$) and congestion-aware ($h=0.95$), respectively. Results are averaged over 100 independent realizations on a scale-free network with $N=1000$ and $\langle k \rangle = 4$.}
\label{fig:3}
\end{figure}

\subsubsection{Global update rule}
\smallskip

Under the global replicator rule, the node-parent results (see Fig.~\ref{fig:3}(a)-(b)) are qualitatively indistinguishable from the packet-centric ones, indicating that the locus of strategy assignment, packet or source node, appears to be secondary to the selection pressure set by the payoffs.

\smallskip

\subsubsection{Local update rules}
\smallskip
As established in Section \ref{sec:NP_description}, the node-parent formalism integrates two distinct local update mechanisms, the Moran process and the Infinite Fermi rule, which differ fundamentally in the way strategic information propagates through the network. Under the Moran process (a stochastic choice), a focal node selects a neighbor as a role model with a probability proportional to that neighbor's payoff, introducing a stochastic element into the adoption decision. Under the Infinite Fermi rule, by contrast, the focal node randomly chooses a neighbor to compare itself to and adopts its strategy only if its payoff is higher than its own. Despite this difference, both local rules (see Fig.~\ref{fig:3}(c)-(f)) reproduce the macroscopic behavior of the global replicator: the network converges to the same $p_c$ and to quantitatively similar partitions $\{f_\alpha\}$ across all three mechanisms. 
\smallskip

The emergent routing efficiency is therefore robust to how strategic information propagates, and does not require global information or centralized coordination since local pairwise interactions suffice.
The only deviation arises under Infinite Fermi with the global-efficiency payoff, where $h=0.5$ attains a markedly higher prevalence at low $p$ than under Moran or the replicator. This follows from the deterministic character of the rule: in the low-$p$ regime the payoffs of $h=0.5$, $0.25$ and $0.05$ are similar but not identical, with $h=0.5$ marginally superior. Moran turns this into only a slightly higher adoption probability, so stochastic exploration prevents dominance\textcolor{red}{. I}nfinite Fermi, by always copying the strictly best neighbor, selects $h=0.5$ with certainty whenever present, letting it spread further.
\smallskip

\begin{figure}[]
\centering\includegraphics[width=0.85\linewidth]{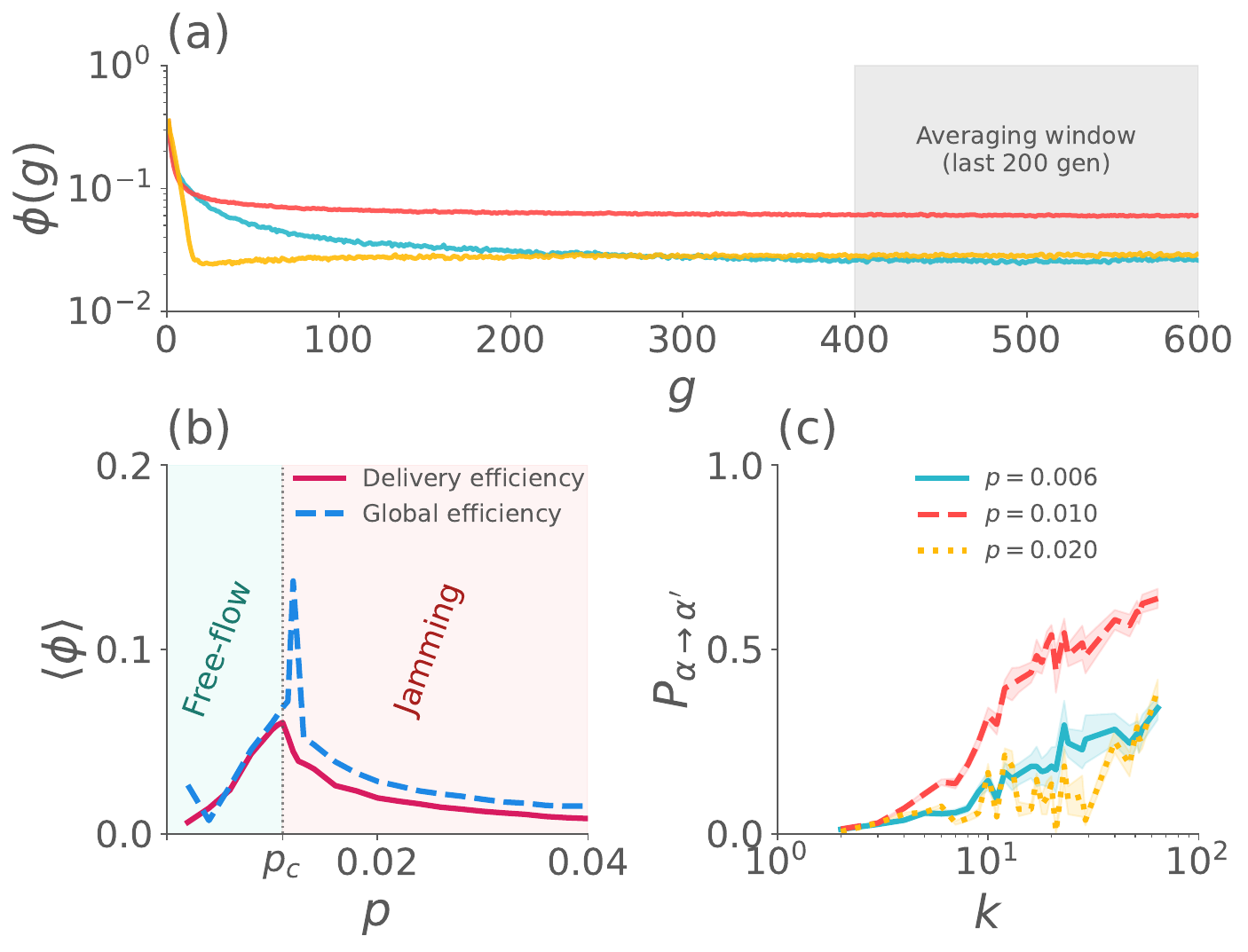}
\caption{\textbf{Locality measurements under Infinite Fermi update rule.} (a) Generational evolution of the mean switching fraction $\phi(g)$, defined as the fraction of nodes that change their assigned routing strategy between consecutive generations $g$ and $g+1$, shown on a semi-logarithmic scale for three representative values of the packet injection probability: $p = 0.006$ (solid blue, sub-critical), $p = 0.01$ (dashed red, near-critical), and $p = 0.02$ (dotted yellow, super-critical). (b) Stationary mean switching fraction $\langle \phi \rangle$⟩, obtained by averaging $\phi(g)$ over the stationary regime ($\sim 200$ generations), plotted as a function of $p$ for the two distinct payoff metrics: delivery-efficiency payoff (solid pink) and global-efficiency payoff (dashed blue). (c) Average probability of changing strategy per generation $P_{\alpha\rightarrow \alpha'}$ as a function of node degree $k$, shown on a semi-logarithmic horizontal scale for the same three values of $p$ as in panel (a), computed under global-efficiency payoff. Results are averaged over 100 independent realizations on a scale-free network with $N=1000$ and $\langle k \rangle = 4$.}
\label{fig:5}
\end{figure}

\subsection{Local early-warning of the jamming transition}
\label{sec:earlywarning}
\smallskip

Because strategies are anchored to nodes, the local update mechanisms grant access to microscopic measurements unavailable in the packet-centric formalism, from which macroscopic signatures of proximity to the jamming transition can be extracted. We introduce the mean switching fraction $\phi(g)$, defined as the fraction of nodes that change their strategy between consecutive generations $g$ and $g+1$. This quantity provides a direct measure of the evolutionary instability of the strategy landscape: high $\phi$ signals a constantly fluctuating population, low $\phi$ an evolutionarily quiescent one in which most nodes have settled.
\smallskip

Panel (a) of Fig.~\ref{fig:5} shows $\phi(g)$ for three representative values of $p$ ($p_1=0.006<p_c$, $p_2=0.01\sim p_c$, $p_3=0.02>p_c$) under Infinite Fermi. In all cases $\phi(g)$ decays rapidly before settling into a stationary fluctuating regime, but its stationary level depends non-monotonically on $p$. At $p_1$ (free flow) and $p_3$ (deep in the congested phase) the switching fraction $\phi$ stabilizes at low values. This pinpoints the emergence of a different but equally stable congested-phase strategy configuration in which local fitness differences have been largely exhausted by the evolutionary dynamics. Instead, the near-critical case $p_2$ sustains a markedly elevated switching fraction, signaling the critical fluctuations that accompany the transition and prevent convergence to a fixed point.
\smallskip

This non-monotonicity dependence of the stationary switching fraction on $p$ is quantified in panel (b) of Fig.~\ref{fig:5}, which plots the generation-averaged $\langle \phi \rangle$ (last 200 generations) against $p$ for both payoffs. Both curves peak sharply near $p_c$, the peak being more pronounced under global-efficiency. Crucially, this peak predicts the jamming transition without measuring $Q(t)$ or $\rho$: $p_c$ can be estimated from the location of the maximum. A network under local evolutionary rules could thus sense the imminence of its own jamming purely from anomalous neighborhood volatility, with no node accessing global statistics.
\smallskip

Panel (c) of Fig.~\ref{fig:5} disaggregates the switching rate by degree, plotting $P_{\alpha\rightarrow \alpha'}$ versus $k$ for the same three values of $p$. Across all conditions the propensity to switch grows monotonically with degree, hub nodes switching far more than peripheral ones. This is natural given the local mechanism and the role of hubs: with large neighborhoods, hubs see a more diverse strategy sample and more intense, fluctuating traffic, generating larger and more variable fitness differentials and hence more frequent adoption. Low-degree nodes, with few and homogeneous neighbors and steadier traffic, settle faster. Although Fig.~\ref{fig:5} only reports Infinite Fermi, this phenomenology is not specific to it: across all node-parent update variants, the mean switching fraction emerges as an accessible local indicator of the dynamical phase, signaling proximity to jamming without any global measure of network state (see Supplementary Material).
\smallskip

Taken together, the node-parent results yield a consistent picture: despite the diversity of update mechanisms, payoff metrics, and local fitness landscapes, the macroscopic congestion behavior is unchanged. The fitness landscape supports multiple local minima whose selection depends on rule, metric, and stochastic trajectory, yet none of these microscopic differences propagate to the global scale. This insensitivity underscores the robustness of the framework and suggests the routing improvement is a generic property of evolutionary strategy competition on scale-free networks, not an artifact of any particular implementation.
\smallskip

\section{Conclusions}
\label{sec:4}

In this work we have presented an evolutionary game theory framework for adaptive packet routing on networks, in which a heterogeneous population of strategies parameterized by a congestion-awareness parameter $h \in [0,1]$ competes for prevalence under selection pressure from differential routing performance. The framework recovers the two fixed benchmarks: pure shortest-path routing undergoes a second-order jamming transition at a low critical rate $p_c^{\mathrm{SP}}$, while congestion-aware routing delays jamming to $p_c^{\mathrm{CA}} > p_c^{\mathrm{SP}}$ at the cost of a sharper, first-order-like collapse, consistent with the literature \cite{Echenique2004PRE, Echenique2005EPL}.
\smallskip

Under all implementations considered (packet-centric and node-parent formalisms, global and local update rules, and both payoff metrics) the evolutionary dynamics yield the same qualitative improvement: the jamming transition is delayed relative to fixed SP routing, and when congestion sets in the system avoids the violent collapse of fixed CA strategies, settling at congestion levels comparable to SP routing. This controlled transition emerges spontaneously, without centralized coordination. Remarkably, it does so despite two features that might \textit{a priori} preclude it: the dynamics are self-interested (with each strategy competing only for its own prevalence and no agent optimizing collective performance) and the myopic exploration of the fitness landscape (with independent runs converging to distinct local configurations). A consistent and reproducible macroscopic outcome emerges across realizations, with the same critical threshold $p_c$ and the same qualitative structure of the stationary strategy distribution. This constitutes a non-trivial instance of self-organization in a complex adaptive system, in which global order arises not from centralized coordination or collective rationality.
\smallskip

A central finding is the robustness of this efficiency across a hierarchy of increasingly realistic informational constraints. The packet-centric formalism, where strategy is an independently assigned attribute of each packet, already captures the essential phenomenology. The node-parent formalism, being closer to real routers that maintain a single persistent forwarding policy since it anchors strategy to the generating node, preserves the same critical thresholds, dominant configurations, and efficiency gains, confirming the results are not an artifact of fine-grained assignment. The most demanding test is the transition to local update rules (Moran and Infinite Fermi), which remove any need for system-wide information since adaptation occurs solely through pairwise interactions between neighboring nodes. That the macroscopic benefits survive even here, with no degradation in $p_c$ or in stationary routing efficiency, shows the emergent robustness is not contingent on global information. This constitutes a finding directly relevant to real infrastructures, where such information is unavailable by definition.
\smallskip

Finally, node-level strategy switching reveals a structural signature of the jamming transition. The mean switching fraction $\langle \phi \rangle$, the fraction of nodes that revise their strategy between consecutive generations, peaks sharply near $p_c$. Crucially, this peak is detectable from purely local observations, without any node accessing global traffic statistics, and thus provides an indicator of proximity to the transition that emerges organically from the dynamics rather than from dedicated monitoring. Disaggregating by degree, high-degree hubs (being exposed to a greater diversity of competing strategies) act as the primary locus of this volatility, with switching rates that grow monotonically with $k$ and peak in the near-critical regime $p \approx p_c$. Taken these results together suggest a form of self-referential congestion awareness, to our knowledge without direct precedent in the adaptive routing literature: a network under local evolutionary rules could exploit anomalous strategic activity as an early-warning signal of imminent jamming~\cite{Scheffer2009Nature}.  Whether this distributed early-warning capacity can be operationalized in practice, for instance, by designing routing agents that modulate their update frequency or strategy exploration range in response to locally observed switching activity, constitutes a compelling direction for future investigation, and one that connects the theoretical framework developed in this work to the applied problem of autonomous congestion management in large-scale communication infrastructures.
\smallskip

\section*{Acknowledgemnet}
F.D., P.G.S and J.G.-G. acknowledge financial support from the Departamento de Industria e Innovaweción del Gobierno de Aragón y Fondo Social Europeo (FENOL group grant E36-23R) and from Ministerio de Ciencia, Innovación y Universidades (grant PID2023-147734NB-I00). F.D. and P.G.S  acknowledge financial support from the Gobierno de Aragón through a doctoral fellowship. S.M. acknowledges financial  support by the Spanish State Research Agency (MICIU/AEI/10.13039/501100011033) and FEDER (UE) under projects COSASTI (PID2024-157493NB-C22), and the Mar{\'\i}a de Maeztu project CEX2021-001164-M.

\bibliographystyle{model1_num_names.bst}

\bibliography{cas_refs}

\newpage

\appendix

\begin{figure}[]
\centering\includegraphics[width=\linewidth]{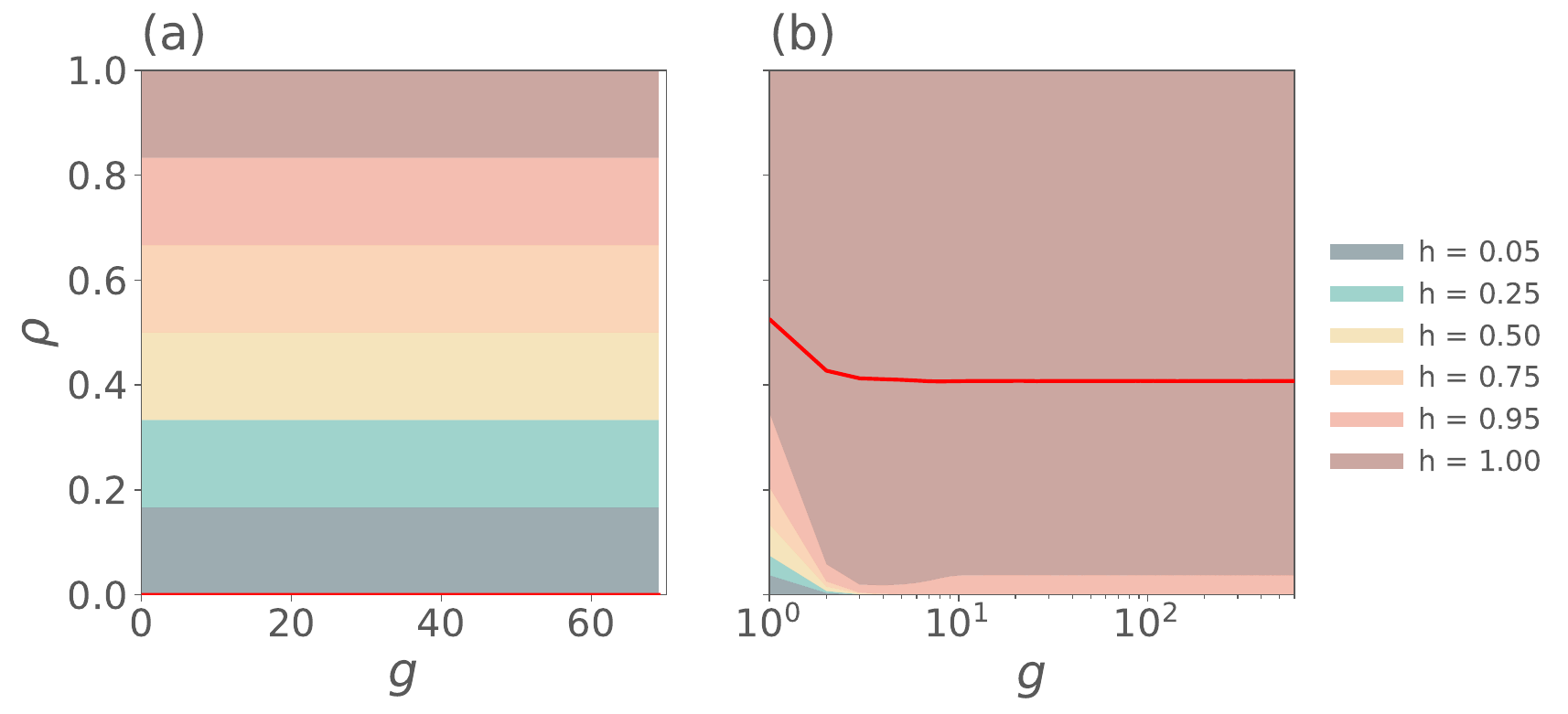}
\caption{\textbf{Generational evolution of the strategy distribution $\{f_\alpha(g)\}$ and $\rho$ under packet-centric formalism and delivery-efficiency payoff $P^{\mathrm{DE}}$.} Generational evolution of the strategy distribution $\{f_\alpha(g)\}$ and the corresponding network congestion level $\rho(g)$ with delivery-efficiency payoff $P^{\mathrm{DE}}$, for two representative values of the packet injection probability: \textbf{(a)} $p = p_1 < p_c$ (free-flow regime) and \textbf{(b)} $p = p_2 > p_c$ (congested regime). Stacked area regions represent the fractional prevalence $f_\alpha(g)$ of each routing strategy $\alpha$, identified by its congestion-awareness 
parameter $h \in \{0.05, 0.25, 0.50, 0.75, 0.95, 1.00\}$ and
color-coded as indicated in the legend. Results are averaged over 100 independent realizations on a scale-free network with $N=1000$ and $\langle k \rangle = 4$.}
\label{fig:SM_1_1}
\end{figure}

\begin{figure}[]
\centering\includegraphics[width=\linewidth]{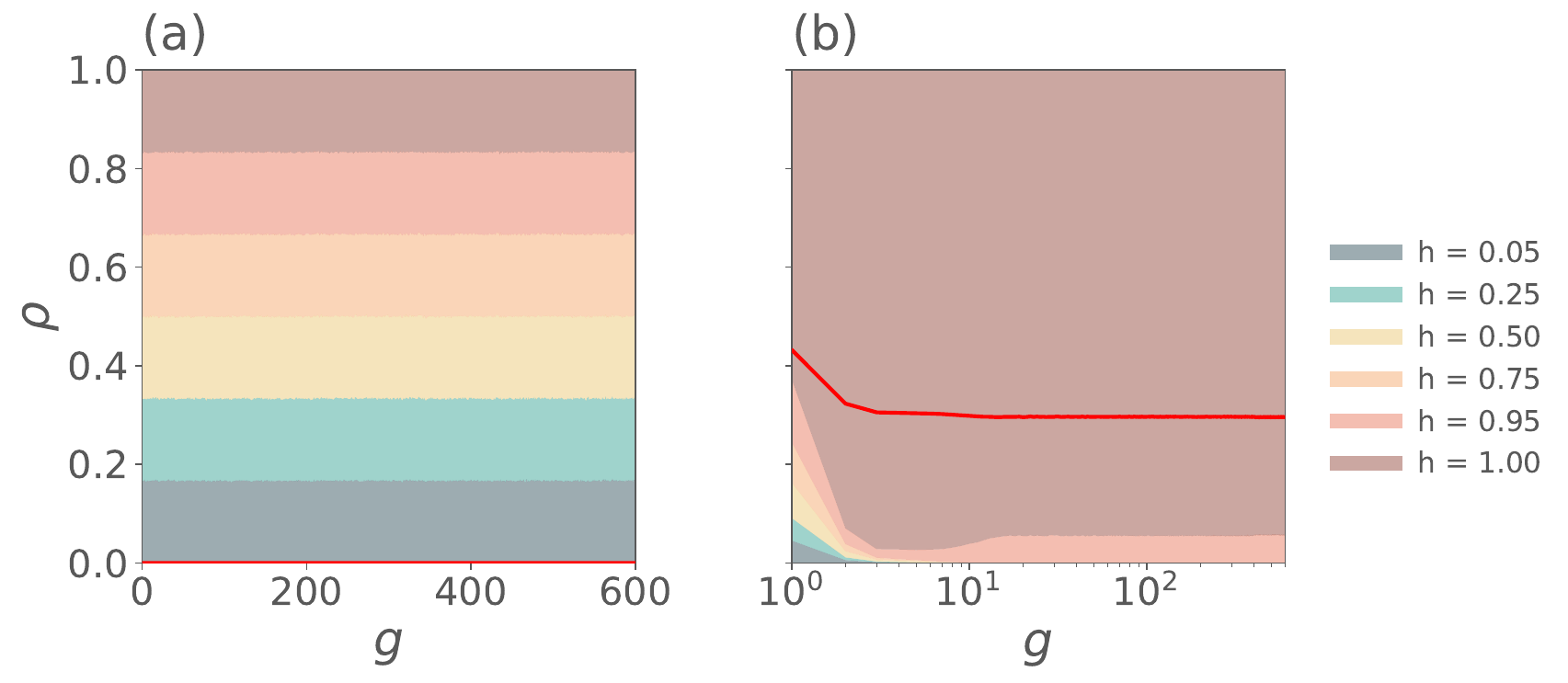}
\caption{\textbf{Generational evolution of the strategy distribution $\{f_\alpha(g)\}$ and $\rho$ under Replicator update rule and delivery-efficiency payoff $P^{\mathrm{DE}}$.} Generational evolution of the strategy distribution $\{f_\alpha(g)\}$ and the corresponding network congestion level $\rho(g)$ under the Replicator update rule ($\beta \to \infty$) with delivery-efficiency payoff $P^{\mathrm{DE}}$, for two representative values of the packet injection probability: \textbf{(a)} $p = p_1 < p_c$ (free-flow regime) and \textbf{(b)} $p = p_2 > p_c$ (congested regime). Stacked area regions represent the fractional prevalence $f_\alpha(g)$ of each routing strategy $\alpha$, identified by its congestion-awareness 
parameter $h \in \{0.05, 0.25, 0.50, 0.75, 0.95, 1.00\}$ and
color-coded as indicated in the legend. Results are averaged over 100 independent realizations on a scale-free network with $N=1000$ and $\langle k \rangle = 4$.}
\label{fig:SM_1_2}
\end{figure}

\begin{figure}[]
\centering\includegraphics[width=\linewidth]{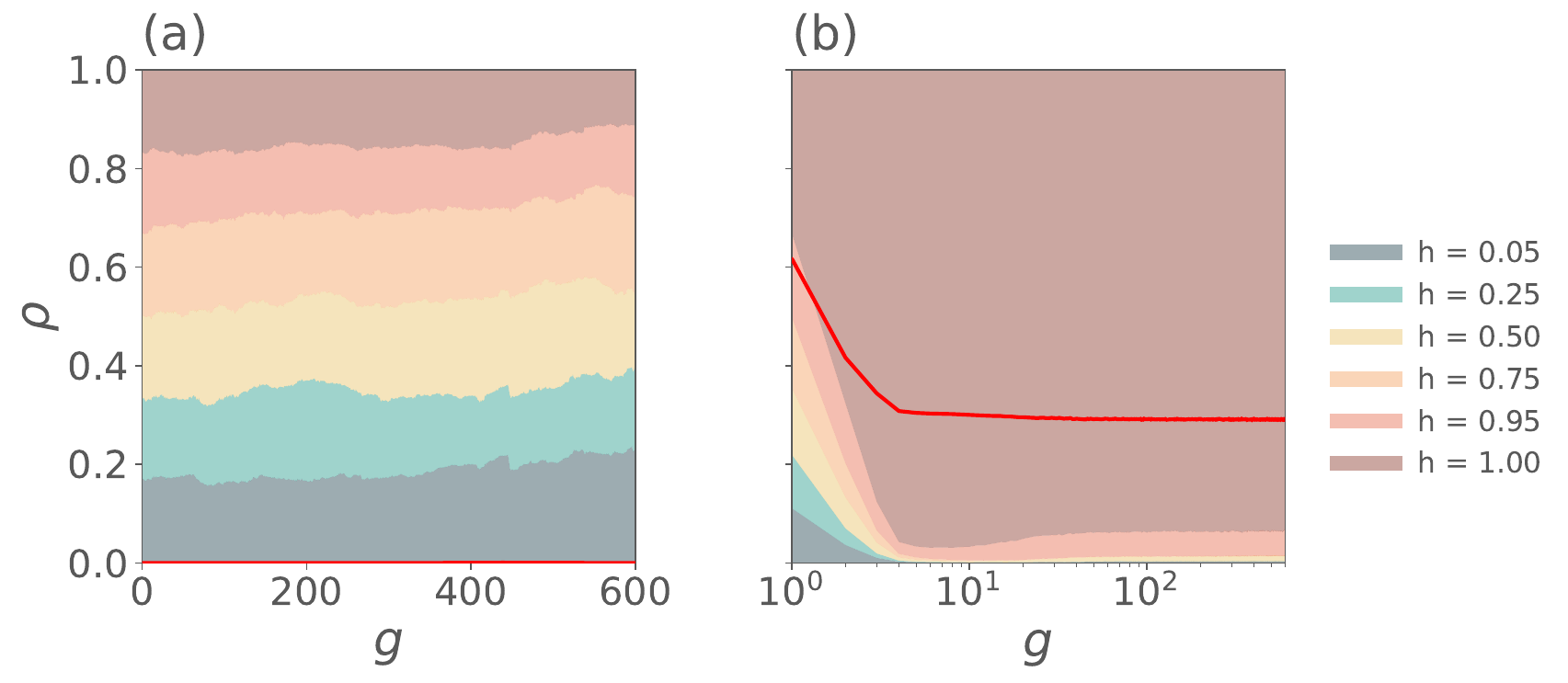}
\caption{\textbf{Generational evolution of the strategy distribution $\{f_\alpha(g)\}$ and $\rho$ under Moran update rule and delivery-efficiency payoff $P^{\mathrm{DE}}$.} Generational evolution of the strategy distribution $\{f_\alpha(g)\}$ and the corresponding network congestion level $\rho(g)$ under the Moran update rule ($\beta \to \infty$) with delivery-efficiency payoff $P^{\mathrm{DE}}$, for two representative values of the packet injection probability: \textbf{(a)} $p = p_1 < p_c$ (free-flow regime) and \textbf{(b)} $p = p_2 > p_c$ (congested regime). Stacked area regions represent the fractional prevalence $f_\alpha(g)$ of each routing strategy $\alpha$, identified by its congestion-awareness 
parameter $h \in \{0.05, 0.25, 0.50, 0.75, 0.95, 1.00\}$ and
color-coded as indicated in the legend. Results are averaged over 100 independent realizations on a scale-free network with $N=1000$ and $\langle k \rangle = 4$.}
\label{fig:SM_1_3}
\end{figure}

\begin{figure}[]
\centering\includegraphics[width=\linewidth]{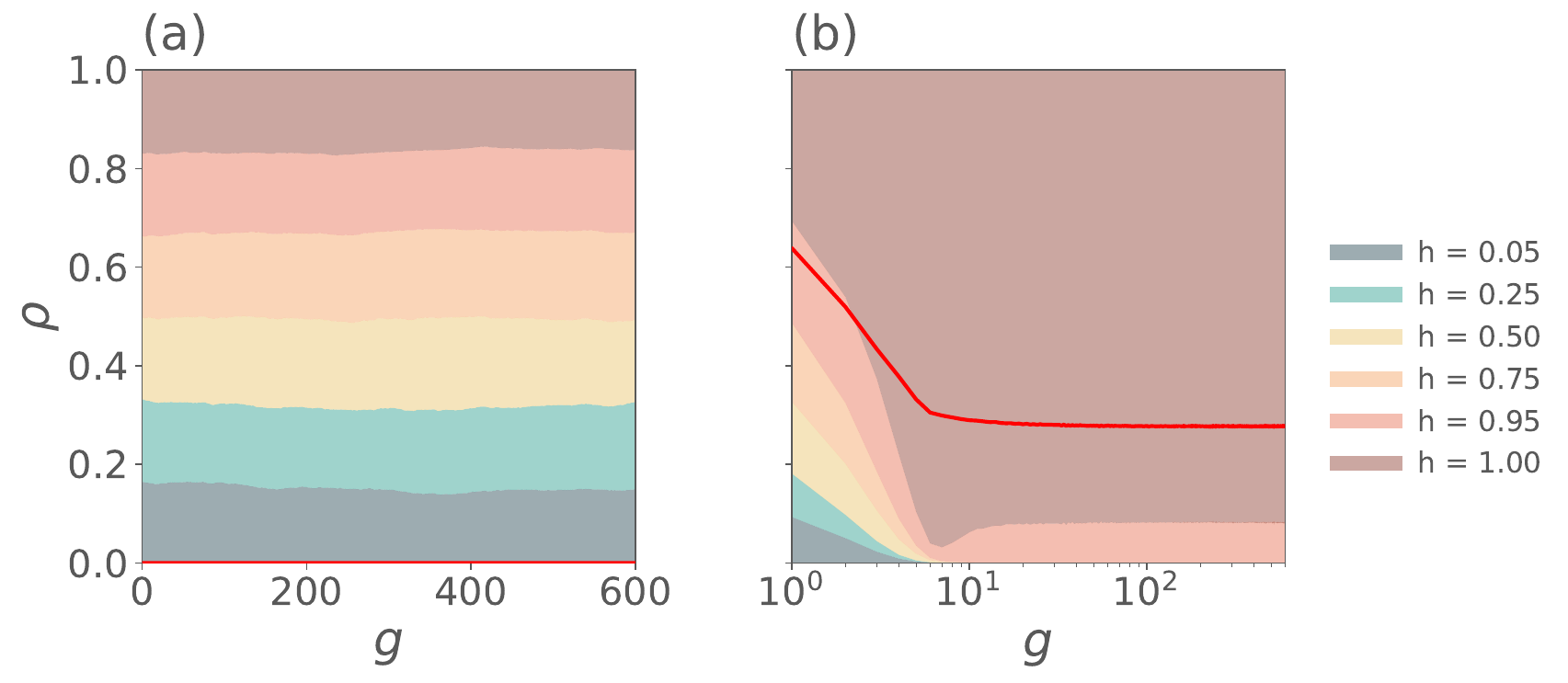}
\caption{\textbf{Generational evolution of the strategy distribution $\{f_\alpha(g)\}$ and $\rho$ under Infinite Fermi update rule and delivery-efficiency payoff $P^{\mathrm{DE}}$.} Generational evolution of the strategy distribution $\{f_\alpha(g)\}$ and the corresponding network congestion level $\rho(g)$ under the Infinite Fermi update rule ($\beta \to \infty$) with delivery-efficiency payoff $P^{\mathrm{DE}}$, for two representative values of the packet injection probability: \textbf{(a)} $p = p_1 < p_c$ (free-flow regime) and \textbf{(b)} $p = p_2 > p_c$ (congested regime). Stacked area regions represent the fractional prevalence $f_\alpha(g)$ of each routing strategy $\alpha$, identified by its congestion-awareness 
parameter $h \in \{0.05, 0.25, 0.50, 0.75, 0.95, 1.00\}$ and
color-coded as indicated in the legend. Results are averaged over 100 independent realizations on a scale-free network with $N=1000$ and $\langle k \rangle = 4$.}
\label{fig:SM_1_4}
\end{figure}

\begin{figure}[]
\centering\includegraphics[width=\linewidth]{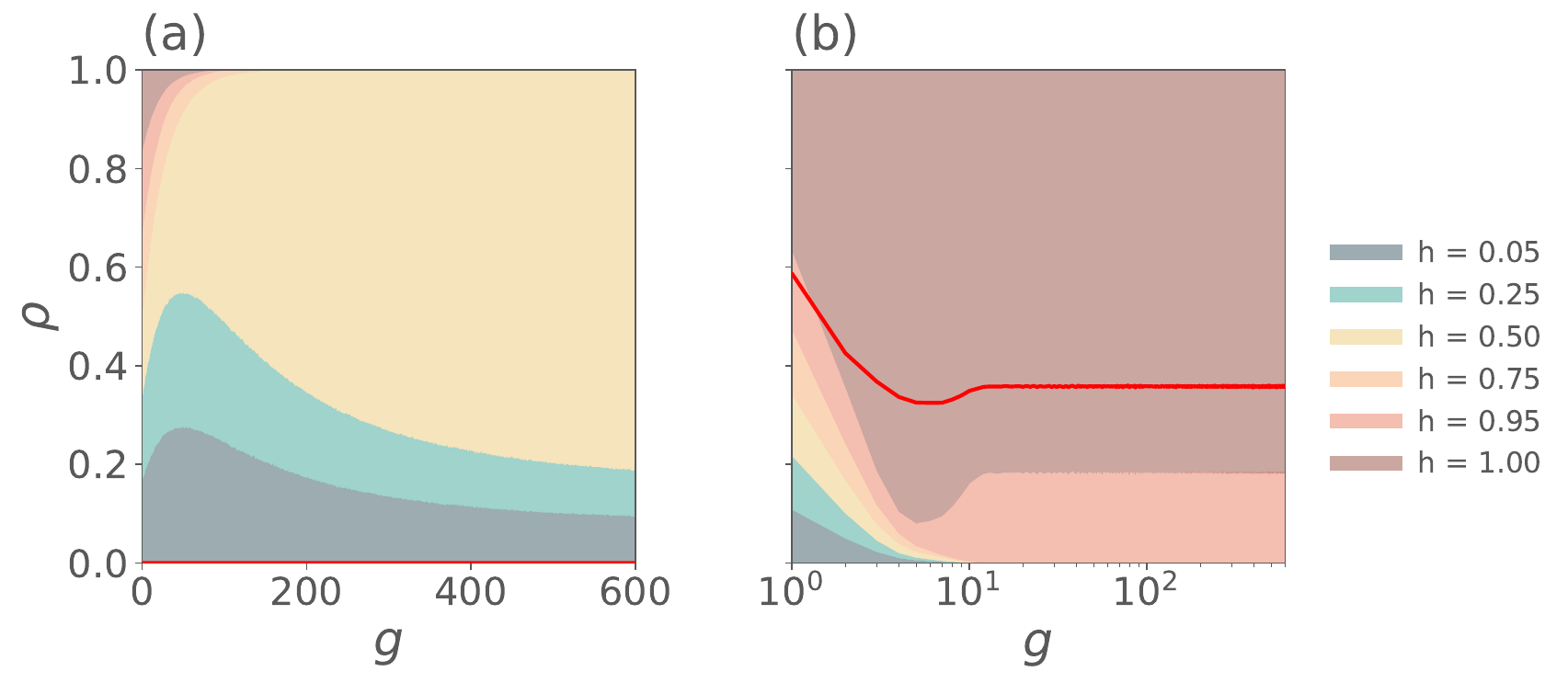}
\caption{\textbf{Generational evolution of the strategy distribution $\{f_\alpha(g)\}$ and $\rho$ under Replicator update rule and global-efficiency payoff $P^{\mathrm{GE}}$.} Generational evolution of the strategy distribution $\{f_\alpha(g)\}$ and the corresponding network congestion level $\rho(g)$ under the Replicator update rule ($\beta \to \infty$) with global-efficiency payoff $P^{\mathrm{DE}}$, for two representative values of the packet injection probability: \textbf{(a)} $p = p_1 < p_c$ (free-flow regime) and \textbf{(b)} $p = p_2 > p_c$ (congested regime). Stacked area regions represent the fractional prevalence $f_\alpha(g)$ of each routing strategy $\alpha$, identified by its congestion-awareness 
parameter $h \in \{0.05, 0.25, 0.50, 0.75, 0.95, 1.00\}$ and
color-coded as indicated in the legend. Results are averaged over 100 independent realizations on a scale-free network with $N=1000$ and $\langle k \rangle = 4$.}
\label{fig:SM_1_5}
\end{figure}

\begin{figure}[]
\centering\includegraphics[width=\linewidth]{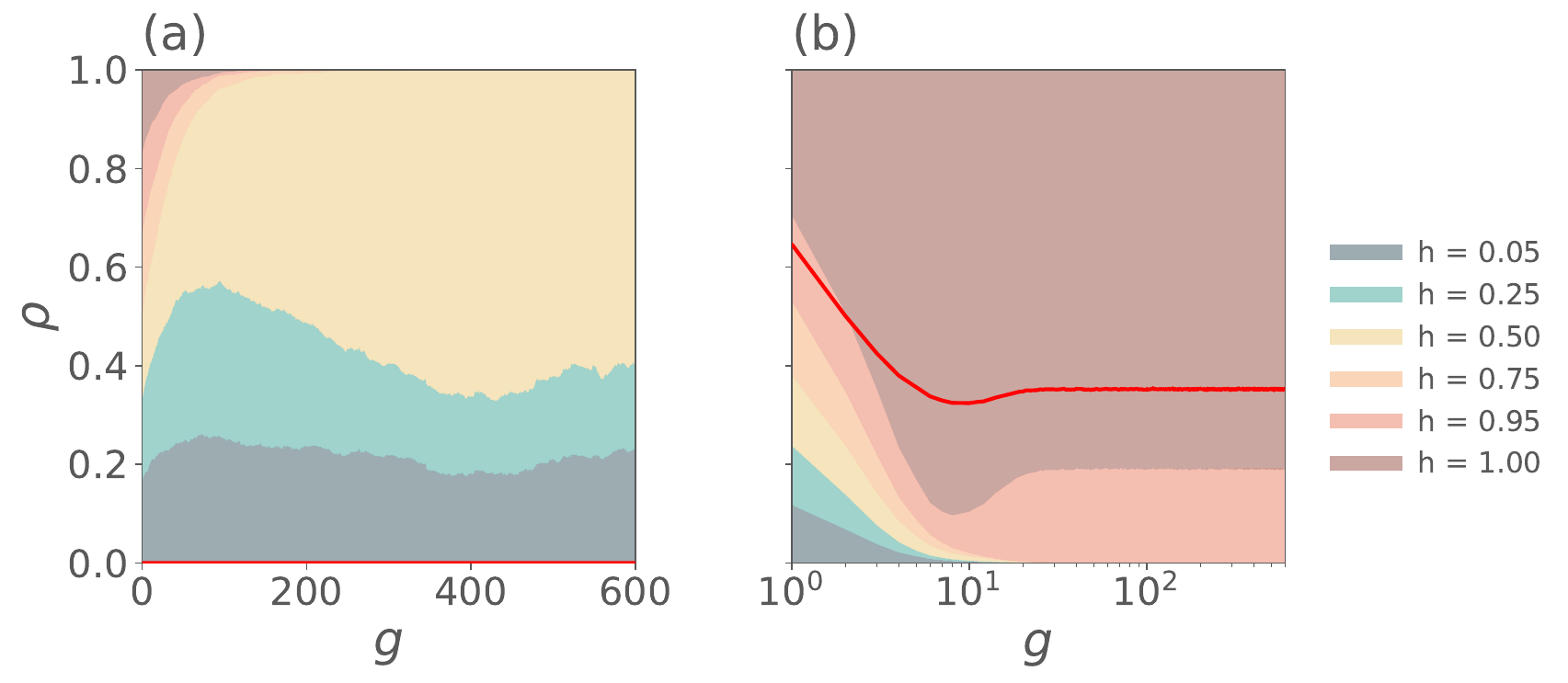}
\caption{\textbf{Generational evolution of the strategy distribution $\{f_\alpha(g)\}$ and $\rho$ under Moran update rule and global-efficiency payoff $P^{\mathrm{GE}}$.} Generational evolution of the strategy distribution $\{f_\alpha(g)\}$ and the corresponding network congestion level $\rho(g)$ under the Moran update rule ($\beta \to \infty$) with global-efficiency payoff $P^{\mathrm{DE}}$, for two representative values of the packet injection probability: \textbf{(a)} $p = p_1 < p_c$ (free-flow regime) and \textbf{(b)} $p = p_2 > p_c$ (congested regime). Stacked area regions represent the fractional prevalence $f_\alpha(g)$ of each routing strategy $\alpha$, identified by its congestion-awareness 
parameter $h \in \{0.05, 0.25, 0.50, 0.75, 0.95, 1.00\}$ and
color-coded as indicated in the legend. Results are averaged over 100 independent realizations on a scale-free network with $N=1000$ and $\langle k \rangle = 4$.}
\label{fig:SM_1_6}
\end{figure}

\begin{figure}[]
\centering\includegraphics[width=\linewidth]{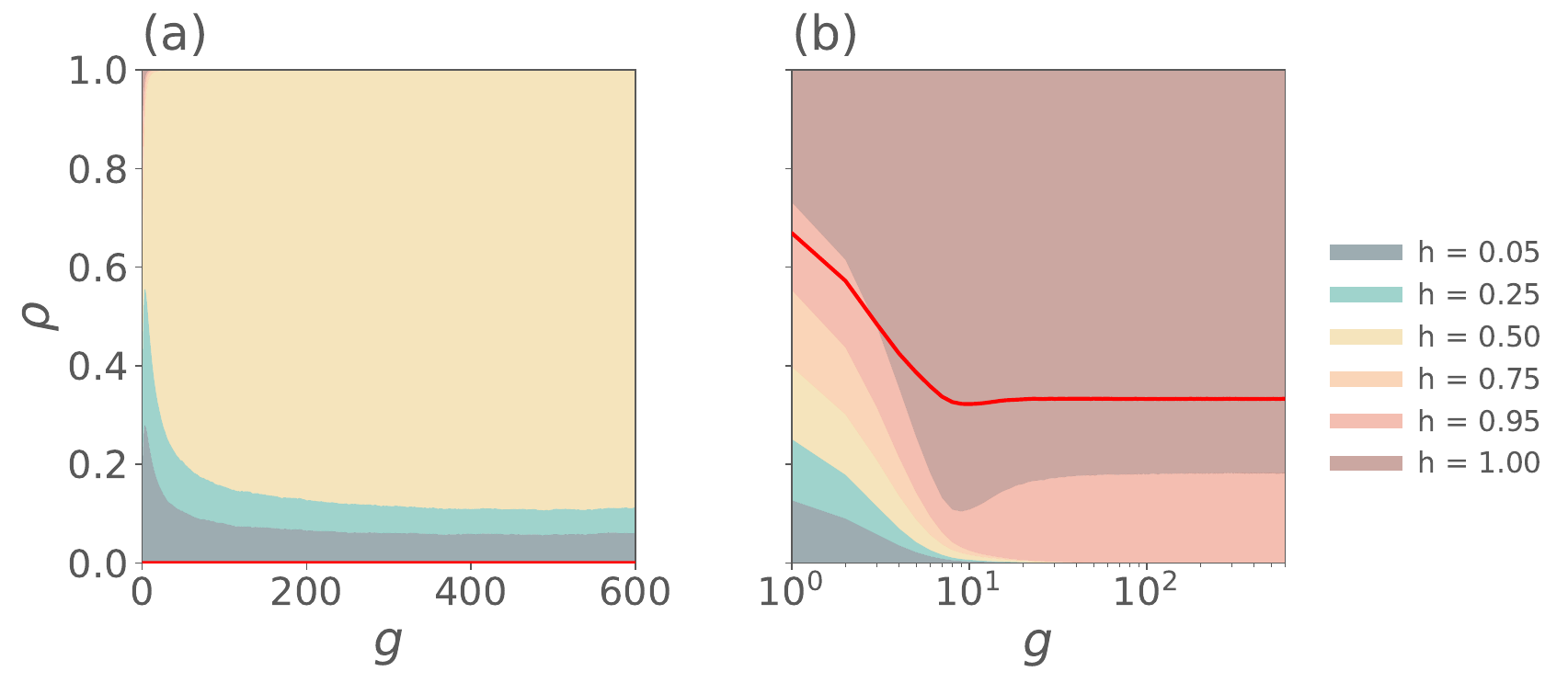}
\caption{\textbf{Generational evolution of the strategy distribution $\{f_\alpha(g)\}$ and $\rho$ under Infinite Fermi update rule and global-efficiency payoff $P^{\mathrm{GE}}$.} Generational evolution of the strategy distribution $\{f_\alpha(g)\}$ and the corresponding network congestion level $\rho(g)$ under the Infinite Fermi update rule ($\beta \to \infty$) with global-efficiency payoff $P^{\mathrm{DE}}$, for two representative values of the packet injection probability: \textbf{(a)} $p = p_1 < p_c$ (free-flow regime) and \textbf{(b)} $p = p_2 > p_c$ (congested regime). Stacked area regions represent the fractional prevalence $f_\alpha(g)$ of each routing strategy $\alpha$, identified by its congestion-awareness 
parameter $h \in \{0.05, 0.25, 0.50, 0.75, 0.95, 1.00\}$ and
color-coded as indicated in the legend. Results are averaged over 100 independent realizations on a scale-free network with $N=1000$ and $\langle k \rangle = 4$.}
\label{fig:SM_1_7}
\end{figure}

\begin{figure}[]
\centering\includegraphics[width=0.45\linewidth]{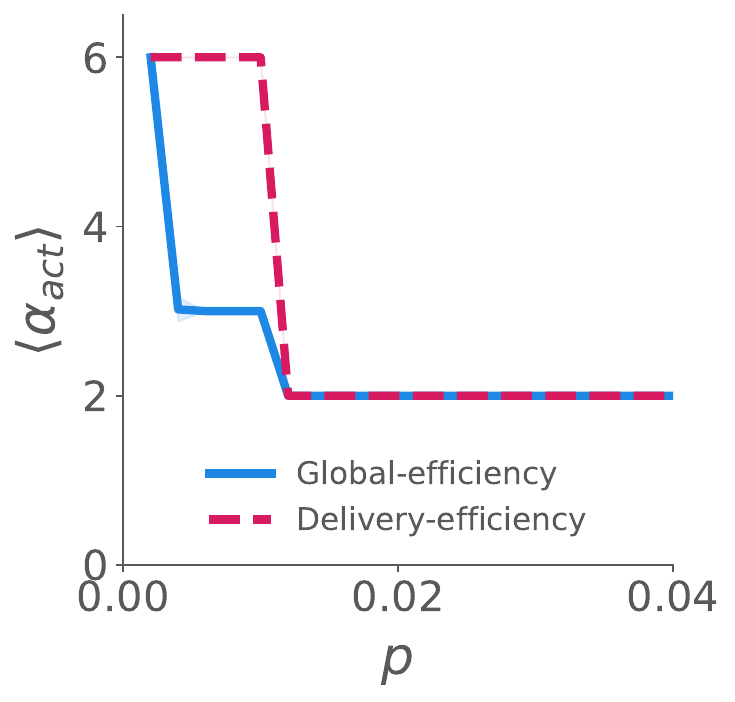}
\caption{\textbf{Mean number of active strategies $\langle \alpha_{\mathrm{act}} \rangle$.} Mean number of active strategies $\langle \alpha_{\mathrm{act}} \rangle$ as a function of the injection probability $p$,  for the packet-centric formalism under three distinct evolutionary update mechanisms, (Replicator (solid cyan), Moran (dashed red), and Infinite Fermi (dotted yellow)) and two payoff metrics: (a) delivery-efficiency payoff $P^{\mathrm{DE}}$ and (b) global-efficiency payoff $P^{\mathrm{GE}}$. A strategy $\alpha$ is counted as active at a given value of $p$ if its fractional prevalence $f_\alpha$ has a non-zero value. Results are averaged over 100 independent realizations on a scale-free network with $N=1000$ and $\langle k \rangle = 4$.}
\label{fig:SM_2}
\end{figure}

\begin{figure}[]
\centering\includegraphics[width=\linewidth]{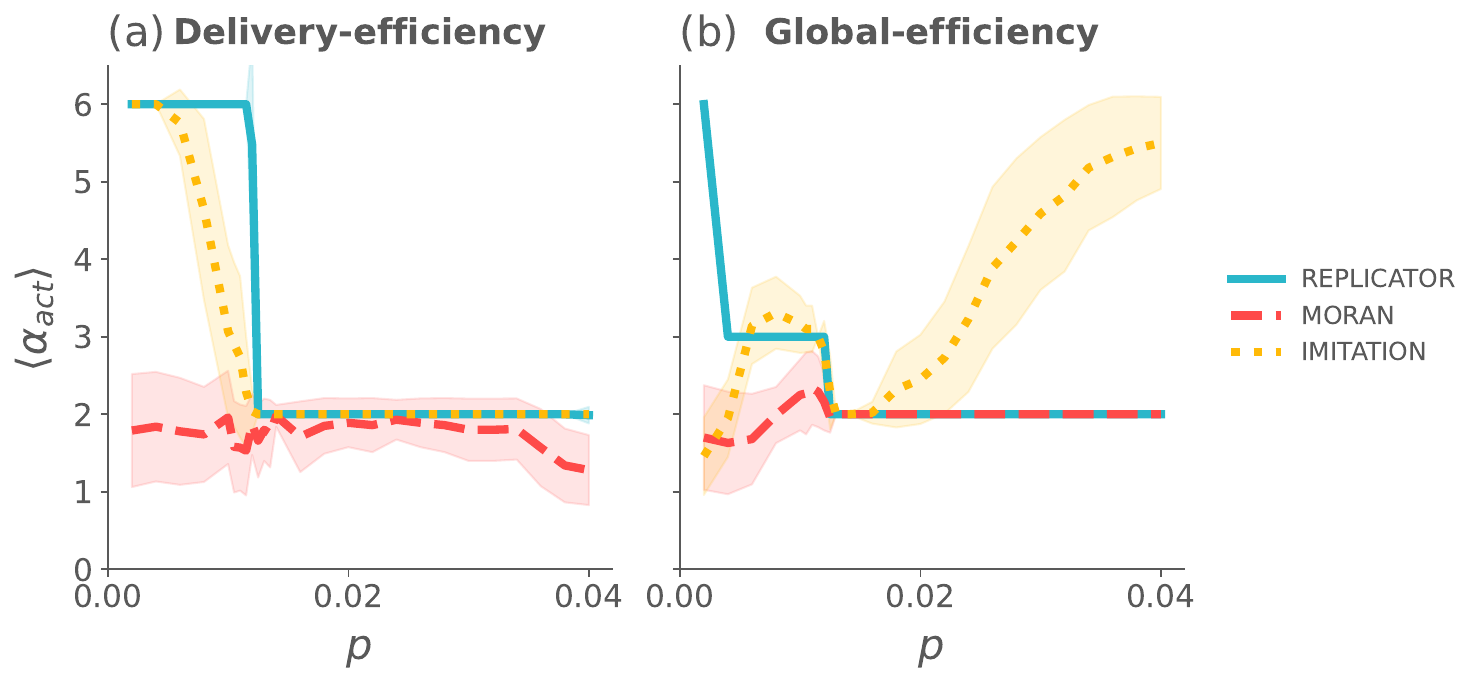}
\caption{\textbf{Mean number of active strategies $\langle \alpha_{\mathrm{act}} \rangle$ for delivery-efficiency and global-efficiency payoffs.} Mean number of active strategies $\langle \alpha_{\mathrm{act}} \rangle$ as a function of the injection probability $p$,  for the node-parent formalism under three distinct evolutionary update mechanisms, (Replicator (solid cyan), Moran (dashed red), and Infinite Fermi (dotted yellow)) and two payoff metrics: (a) delivery-efficiency payoff $P^{\mathrm{DE}}$ and (b) global-efficiency payoff $P^{\mathrm{GE}}$. A strategy $\alpha$ is counted as active at a given value of $p$ if its fractional prevalence $f_\alpha$ has a non-zero value. Results are averaged over 100 independent realizations on a scale-free network with $N=1000$ and $\langle k \rangle = 4$.}
\label{fig:SM_3}
\end{figure}

\begin{figure}[]
\centering\includegraphics[width=\linewidth]{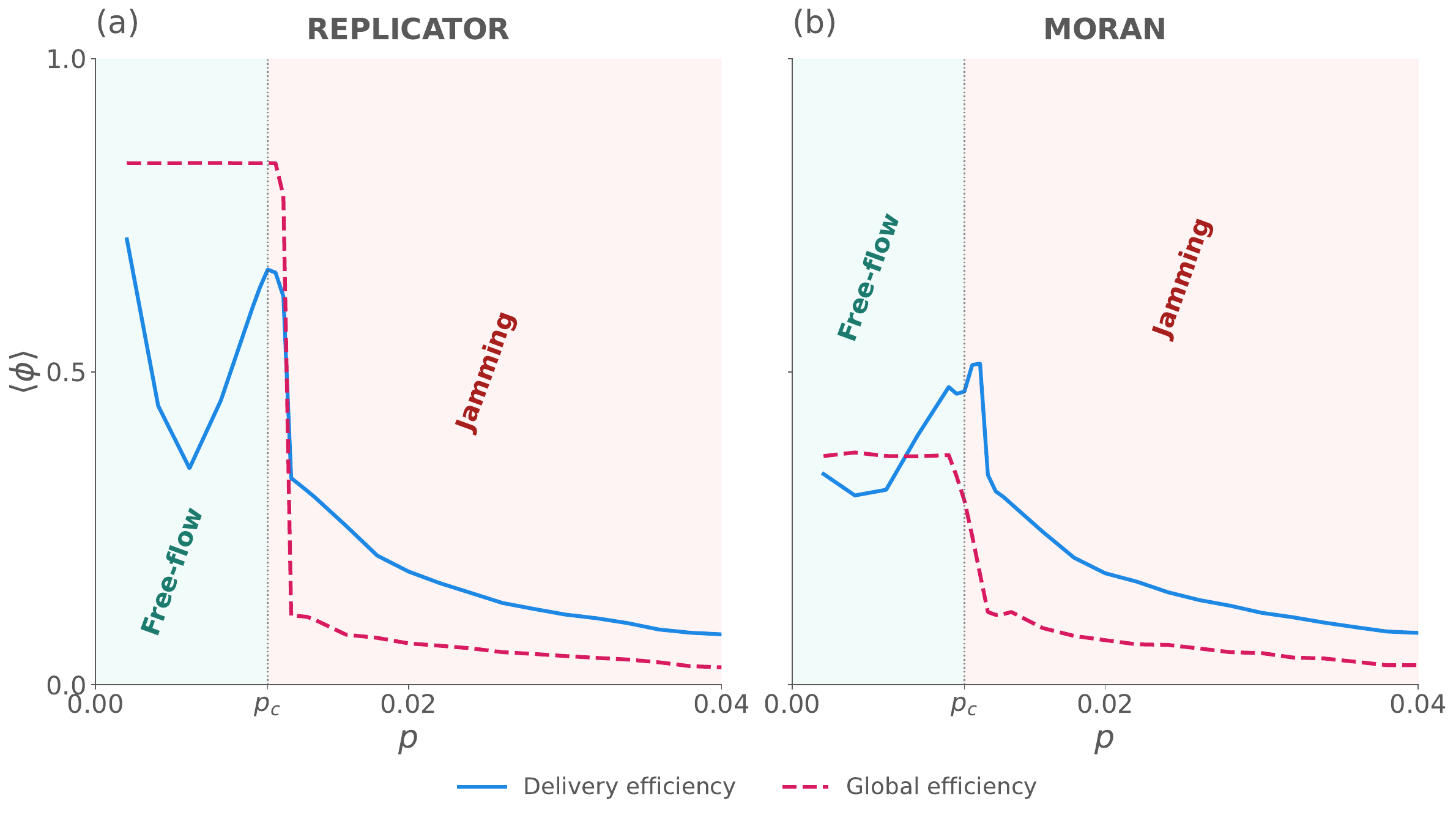}
\caption{\textbf{Stationary mean switching fraction $\langle\phi\rangle$ under Replicator and Moran update rules.} Stationary mean switching fraction $\langle \phi \rangle$⟩, obtained by averaging $\phi(g)$ over the stationary regime ($\sim 100$ generations), plotted as a function of $p$ for the two distinct payoff metrics: delivery-efficiency payoff (solid pink) and global-efficiency payoff (dashed blue) and under two update rules: (a) Replicator and (b) Moran. Results are averaged over 100 independent realizations on a scale-free network with $N=1000$ and $\langle k \rangle = 4$.}
\label{fig:SM_4}
\end{figure}

\end{document}